\documentclass[12pt]{article}
\usepackage{ifpdf}
\usepackage{graphicx,subfigure}
\usepackage[english]{babel}
\usepackage{mathrsfs}
\usepackage{amsfonts, amssymb}

\ifx\pdfoutput\undefined
\usepackage[hypertex]{hyperref}
\else
\usepackage[pdftex,hypertexnames=false]{hyperref}
\fi

\begin{document}
\thispagestyle{empty}
\vfill
\begin{center} {\Large Bernstein-based polynomial approach to study the stability of switched systems and formal verification using HOL Light} \end{center}
\vfill
\begin{center}{\sc Lo\"ic MICHEL} \end{center}

\begin{center}
\end{center}
\vfill

\begin{quotation}                
\begin{center} {\bf Abstract} \end{center}
\noindent
In this preliminary work\footnote{This work is distributed under CC license \url{http://creativecommons.org/licenses/by/3.0/}. Email of the corresponding author : loic.michel54@gmail.com},
we propose to use a polynomial approach 
in order to study the stability of switched systems. The proposed strategy is based on the Bernstein interpolation method that may transform a switched system into a polynomial
expression from which an associated "simple" Lyapunov function can be eventually built. The HOL Light proof assistant allows verifying formally 
the Lyapunov functions that are identified from the proposed switching structure. Our approach is illustrated by numerical examples.
\end{quotation}
\vfill
\vfill
\def\thefootnote{\fnsymbol{footnote}}
\setcounter{footnote}{0}

\newpage
\setcounter{page}{1}
\section{Introduction}
A lot of investigations have been focusing on switched systems (e.g. \cite{Liber} \cite{Liber2} \cite{Lin} \cite{Shorten} \cite{Eben}) and many successful methods have 
been proposed to conclude on the stability of switched systems. 
In this brief, we develop a polynomial expression that describes the full behavior of a switched system, composed of several "switched" 
subsystems that follow a switching signal. Based on the work of \cite{Mojica}, we propose to transpose the switching signal into a polynomial formulation that "connects" each 
subsystem. The proposed method used to reformulate the switching signal is based on a Bernstein interpolation that allows to build a Lyapunov candidate function. We propose to use the formal verification of nonlinear inequalities methodology to check the properties of such Lyapunov candidate function. 
This paper is structured as follow. Section II presents the outline of the proposed method including a brief presentation of the Bernstein interpolation procedure. Section III presents
some simulation results. Section IV presents a general methodology to check properties of Lyapunov candidate functions in order to conclude about stability of dynamical systems.
Some concluding remark may be found in Section V.

\section{Outline of the method}

Following the definition given in \cite{Liber}, a switched system is described by a family $f_p, p \in \mathcal{P}$ of functions from $\mathbb{R}^n$ to $\mathbb{R}^n$ where 
$\mathcal{P}$ is some index set. The variable $p$ is driven by a switching signal $\sigma$, whose role is to specify, according to the switching conditions (for example, $\sigma$ can 
depend on the time $t$), the corresponding $p$ function which is active. A switched system can be described by the equation:

\begin{equation}\label{eq:base}
\frac{d x(t)}{d t} = f_{\sigma(t)} x(t)
\end{equation}

\noindent
where the elements $f_p$ are also called subsystems of the switched system $f$ ; $\sigma$ is the switching signal that takes the values $\sigma_1, \sigma_2, \cdots, \sigma_n$ according to 
the conditions of switching that drive the switching signal.
We assume, in this paper, that $f$ is linear and to define a unified expression of (\ref{eq:base}), we start from the complete description of (\ref{eq:base}):

\begin{equation}\label{eq:base_exp}
\left\{ \begin{array}{l}
 \displaystyle{  \frac{d x(t)}{d t} = f_1 (x(t)) }  \, \hbox{ if } \, \sigma_1 \\      
 \displaystyle{  \frac{d x(t)}{d t} = f_2 (x(t)) }  \, \hbox{ if } \, \sigma_2 \\ 
 \vdots \\
 \displaystyle{  \frac{d x(t)}{d t} = f_n (x(t)) }  \, \hbox{ if } \, \sigma_n \\
        \end{array} \right.
\end{equation}

\noindent
As a result, (\ref{eq:base_exp}) is rewritten as a convex combination of $f_1, f_2, \cdots, f_n$ :

\begin{equation}\label{eq:base_convex}
  \frac{d x(t)}{d t} = \sum_{i = 1}^n  f_i(x(t)) \sigma_i \quad \hbox{ with }  \sum_{i = 1}^n \sigma_i = 1  \quad \forall i, \sigma_i \geq 0
\end{equation}

\noindent
To obtain an expression of (\ref{eq:base_convex}), which is completely written in terms of polynomials, we approximate the switching signal $\sigma$ by a polynomial expression 
based on the Bernstein interpolation.

\paragraph{Bernstein polynomial-based interpolation methodology}

The Bernstein polynomials of degree $m  \in \mathbb{N}^{*+}$ are defined by :

\begin{equation}
 \mathcal{B}_{ m} (t) = \left( \begin{array}{c}
                       m \\
                       i \\
                      \end{array} \right) t^i (1 - t)^{n-i} 
\end{equation}

\noindent
for all $i \in \mathbb{N}^+$, where :

\begin{equation} 
\left( \begin{array}{c}
  m \\
  i \\
 \end{array} \right) = \frac{m !}{i ! ( m - i) !}              
\end{equation}

\noindent
It is well-known (e.g. \cite{Philips}) that, given a function $f$ on $[0,\, 1]$, the associated Bernstein polynomial reads:

\noindent
\begin{equation}\label{eq:bernstein_exp}
 \mathcal{B}_m(f,x) = \sum_{r = 0}^m f \left( \frac{r}{m} \right) \left( \begin{array}{c}
  m \\
  r \\
 \end{array} \right) x^r (1 - x)^{m - r}
\end{equation}

\noindent
It can be proved that if $f$ is continuous on $[0, \, 1]$, then the sequence of Bernstein polynomials converges uniformly to $f$ on $[0, \, 1]$.
\noindent
We call {\it Bernstein series} of the function $f$ the sequence of Bernstein polynomials that interpolates the function $f$.

\paragraph{Bernstein series of the switching signal}

The purpose is to interpolate the behavior of the switching signal depending on its properties (e.g. state-dependent, time-dependent). 
Assuming that $\sigma$ depends on a generic variable $\nu$, using the Bernstein expansion (\ref{eq:bernstein_exp}), (\ref{eq:base_convex}) can be rewritten:

\begin{equation}\label{eq:new_1}
  \frac{d x(t)}{d t} = \sum_{i = 1}^n \mathcal{B}^{\sigma}_{i, m}(\sigma(\nu), \nu) \, x(t); \quad \hbox{ with }  \sum_{i = 1}^n \mathcal{B}^{\sigma}_{i,m} = 1;  \quad 
  \forall i, \mathcal{B}^{\sigma}_{i,m} \geq 0
\end{equation}

\noindent
where $\mathcal{B}^{\sigma}_{i, m} \equiv \sigma_i$ is the switching signal that is a linear combination of the Bernstein interpolation polynomials.
\noindent
We call (\ref{eq:new_1}) a Bernstein representation of the switched system (or simply a B-switched system) for which we assume that the convergence behavior shares the same 
properties with the "standard" switched system (\ref{eq:base}). In particular, {\it we assume that the stability property of the B-switched system is equivalent to the stability property 
of the corresponding "standard" switched system (at least for the Lyapunov stability property)}.

\paragraph{State-dependent switching signal} The state-dependent case is a sign function that activates the considered subsystem according generally to the sign of the product of the states. 
 For example, consider two subsystems $S_1$ and $S_2$ with two states $x_1$ and $x_2$ with $m = 100$ for which the switching signal 
 is defined by the function $\sigma \equiv \hbox{sign}(\delta x)$ where $\delta \in \mathbb{R}^{*+}$ is a scaling factor. The corresponding $\mathcal{B}^{\sigma}$ signals read:
 
 \begin{equation}\label{eq:b_sigma_def}
 \mathcal{B}^{\sigma}_{1, 100} = \mathcal{B}_{100} ( \mathrm{sign} (x),x); \qquad \mathcal{B}^{\sigma}_{2, 100} = 1 - \mathcal{B}_{100} ( \mathrm{sign} (x),x); \quad x = \delta x_1 x_2
 \end{equation}
 
\begin{figure}[!h]
\centering
\includegraphics[width=15cm]{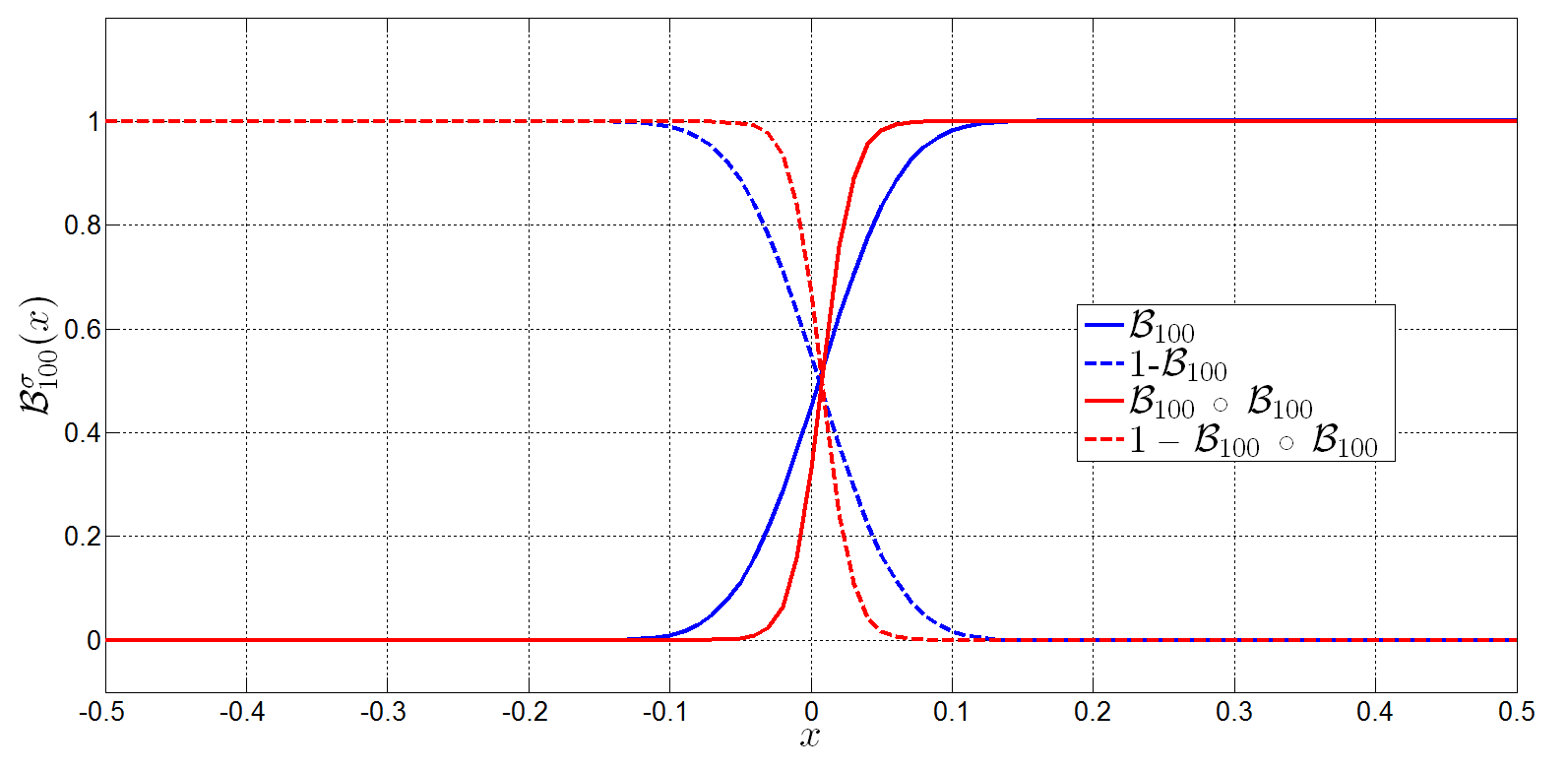}
\caption{Bernstein interpolation of the sign function that describes the switching signal $\sigma = \mathcal{B}_{100}(\hbox{sign}(x),x)$ with $x = \delta x_1 x_2 $.}
\label{fig:B_atan}
\end{figure}

 The corresponding Bernstein interpolation gives the representation depicted in Fig. \ref{fig:B_atan}. 
 In order to better approximate the sign function, one may consider refining the Bernstein approximation by adjusting the value of the integer $m$, or by using the composition 
 rule such that symbolically $\mathcal{B}^{\sigma}_{100} \equiv \mathcal{B}_{100} \circ \mathcal{B}_{100}$. Figure \ref{fig:B_atan} presents also the graphical comparison between the different approximations.
 Although such composition rule gives a more precise practical definition of the B-switched system, an eventual corresponding Lyapunov function would have therefore a more complex 
 expression.

In the particular case of two switching signals, a single Bernstein function can therefore describe the full behavior of the switched system ($S_1, S_2$). 
\begin{figure}[!h]
\centering
\includegraphics[width=16cm]{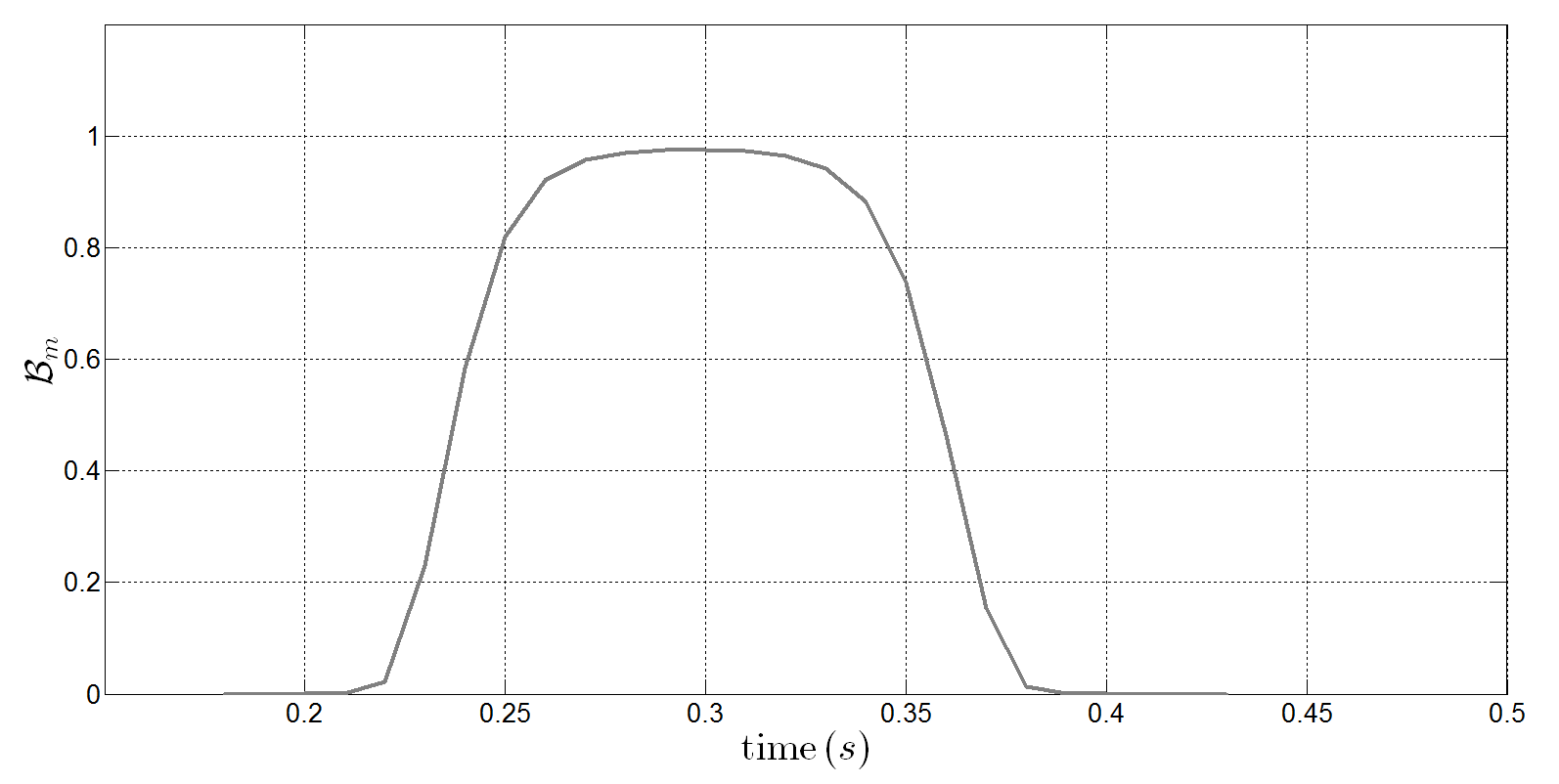}
\caption{Bernstein interpolation of a pulse function that describes the active switching signal $\sigma$ in the time interval $[0.2, \, 0.4]$.}
\label{fig:B_pulse}
\end{figure}
\begin{figure}[!h]
\centering
\includegraphics[width=16cm]{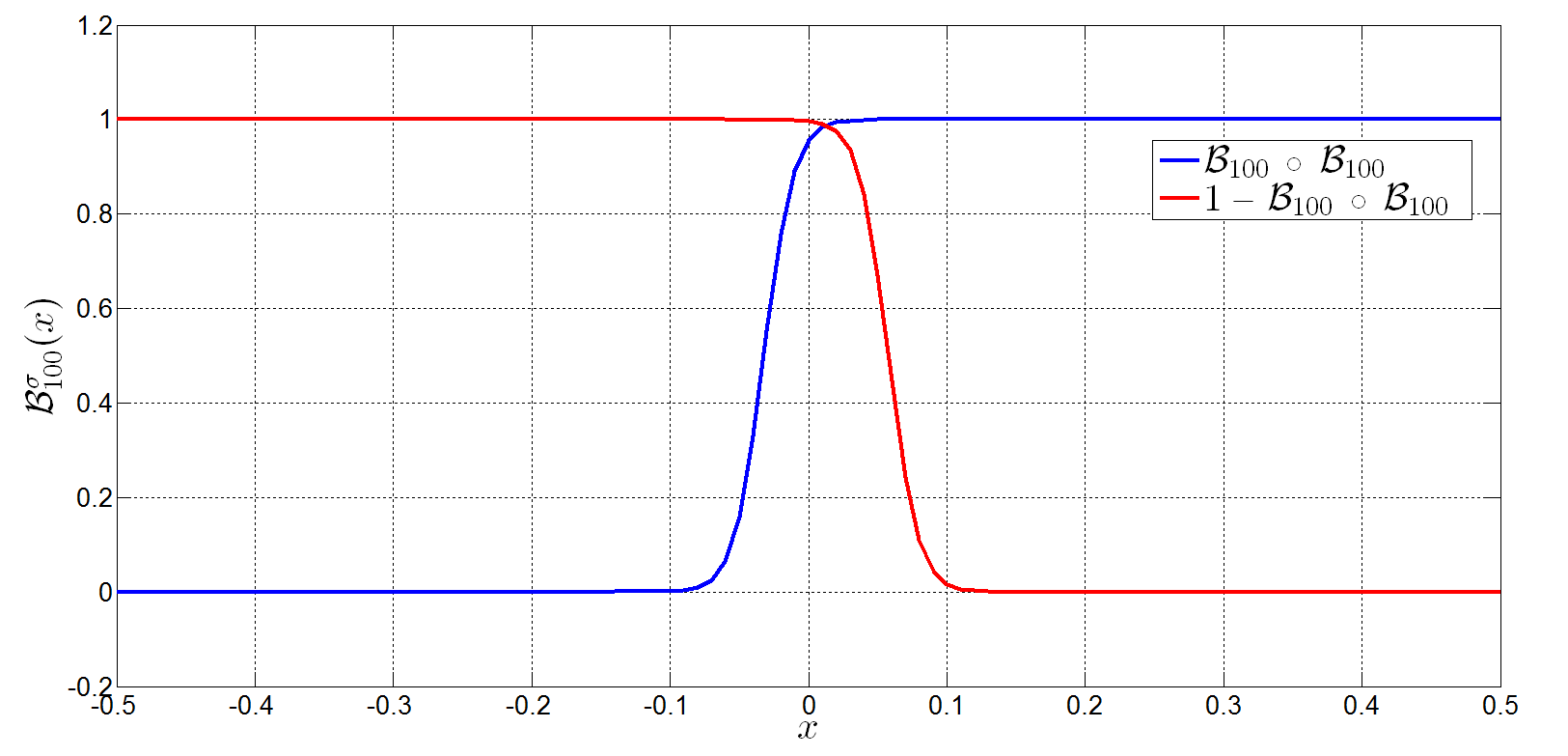}
\caption{Other B-interpolations of the sign function that describe the switching signal.}
\label{fig:biB_shape}
\end{figure}

\paragraph{Remark 1 :} 
The time-dependent case could be considered with a pulse function that activates the considered subsystem according to a specific period of time. For example, consider two subsystems 
 $S_1$ and $S_2$ with two states $x_1$ and $x_2$ for which the switching signal periodically activates alternatively $S_1$ and $S_2$. The corresponding Bernstein interpolation 
 of such pulse gives the representation depicted in Fig. \ref{fig:B_pulse}.
 
\noindent
In this particular case, multiple Bernstein functions are required to create the periodicity of the activation \footnote{Since the Bernstein interpolation is only valid in 
the interval $[0, 1]$, some improved strategies have to be defined in order to deal with infinite simulation horizon.}. A mix of state-dependent and time-dependent switching 
signals can be defined in order to achieve all possible behaviors of the global switched system.
\paragraph{Remark 2 :} The flexibility of the Bernstein expansion (also B-interpolation) allows "shaping" the switching signal such that different "transients" between the subsystems 
can be considered (Fig. \ref{fig:biB_shape}).

\section{Illustrative examples}

\begin{figure}[!h]
\centering
\includegraphics[width=16cm]{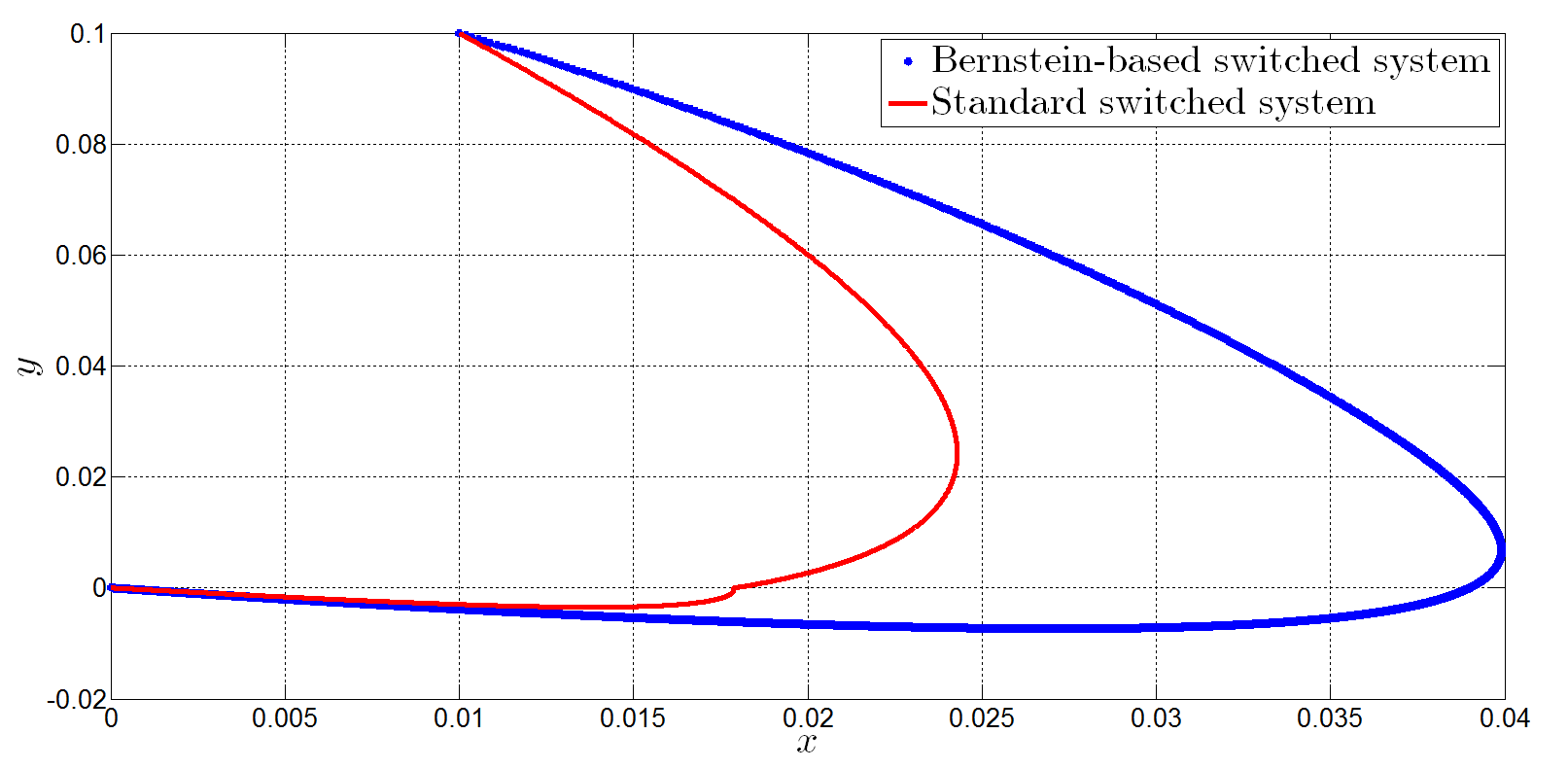}
\caption{Comparison of the B-switched system and the standard system in the case of a stabilizing switching signal.}
\label{fig:stable_sys}
\end{figure}

\begin{figure}[!h]
\centering
\includegraphics[width=16cm]{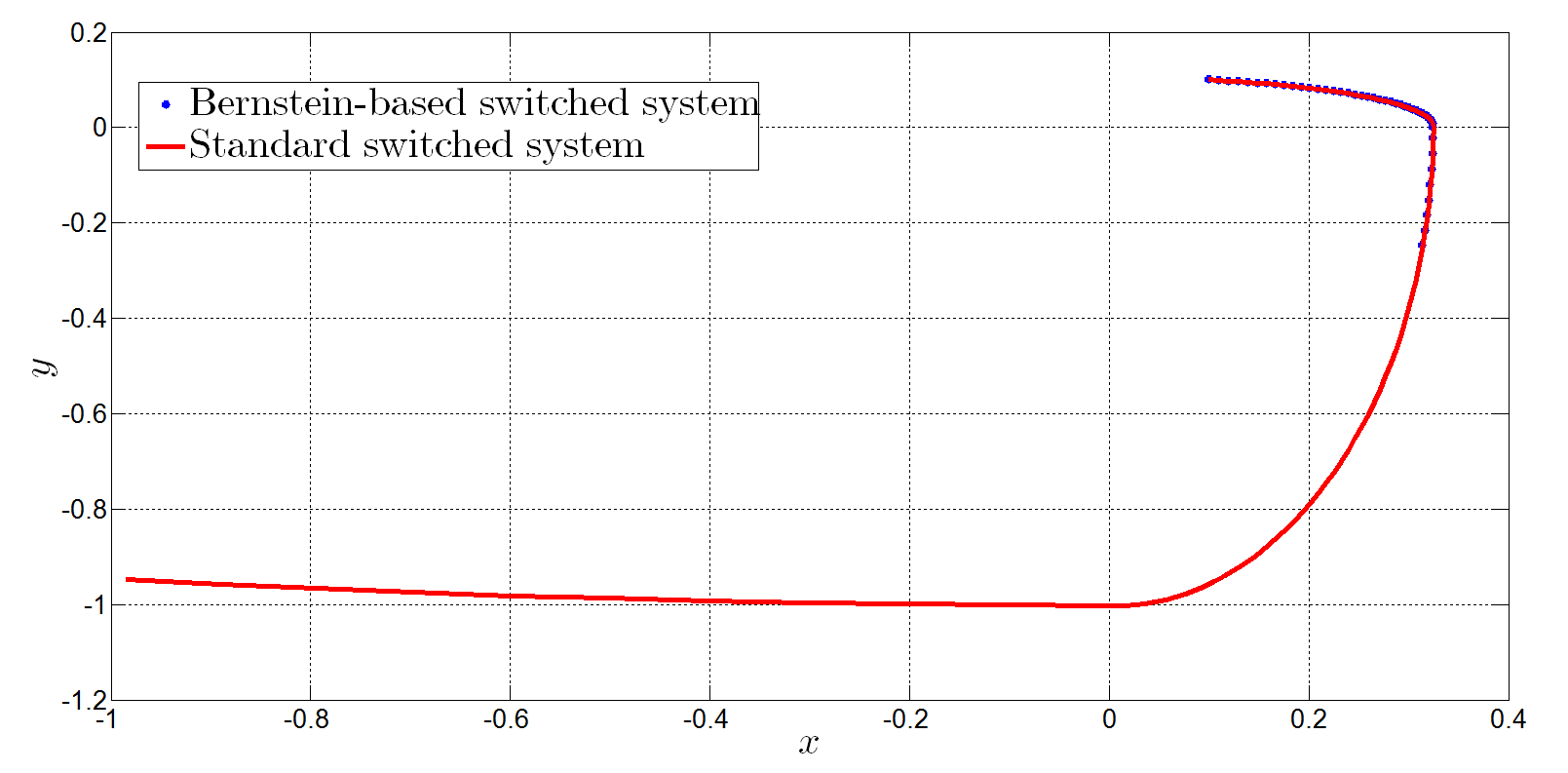}
\caption{Comparison of the B-switched system and the standard system in the case of a non stabilizing switching signal.}
\label{fig:unstable_sys}
\end{figure}

To illustrate the proposed approach, we present two examples, taken from \cite{Bourdais}, that simulate the behavior of a B-switched system in comparison with a "standard" switched 
system. Consider two subsystems $S_1$ and $S_2$ with two states $x_1$ and $x_2$ with $m = 100$, for which the switching signal is defined by \ref{eq:b_sigma_def}.
\paragraph{Example 1 :} (p. 54) Consider the case of a stable linear switched system (Fig. \ref{fig:stable_sys} ), composed of two subsystems 
$f_i(x) = A_i x, \, i \in \{1, 2\}$ with :
\begin{equation}
 A_1 = \left( \begin{array}{cc}
 -1 & 1 \\
 -1 & -3 \\
 \end{array} \right); \qquad
  A_2 = \left( \begin{array}{cc}
 0.01 & 3 \\
 -1 & -4\\
 \end{array} \right)
\end{equation}
\paragraph{Example 2 :} (p. 43) Consider the case of an unstable linear switched system (Fig. \ref{fig:unstable_sys}), composed of two subsystems 
$f_i(x) = A_i x, \, i \in \{1, 2\}$ with :
\begin{equation}
 A_1 = \left( \begin{array}{cc}
 -1 & 10 \\
 -100 & -1 \\
 \end{array} \right); \qquad
  A_2 = \left( \begin{array}{cc}
 -1 & 100 \\
 -10 & -1\\
 \end{array} \right)
\end{equation}
\paragraph{Remark 3 :} In the case of a divergent switched system, the product $x_1 x_2$ will naturally increase and be greater than 1. Therefore, the corresponding Bernstein series 
will be also divergent (see Fig. \ref{fig:unstable_sys}).

\newpage
\section{Formal verification of Lyapunov functions}
From the B-switched system (the B-switched system describes a switched system with a "single" polynomial expression) that can be thus considered as a "classical" nonlinear state-space representation
(according to (\ref{eq:new_1})), it is therefore possible to search a Lyapunov candidate function. Once a candidate function has been identified, the tool 
for formal verification of nonlinear inequalities \cite{solo}, developed in the framework of the successful Flyspeck project \cite{flys} and implemented in the HOL Light proof assistant 
\cite{hol}, can be used to verify the properties of the considered Lyapunov candidate function and conclude eventually about the stability of dynamical systems. A similar methodology has been successfully
implemented in \cite{Why}.
\subsection{General considerations}
We consider the method of interval arithmetic with Taylor approximations as a very efficient way to verify nonlinear inequalities \cite{solo}. 

Consider the problem :
\begin{equation}\label{eq:main_taylor}
 \forall \mathbf{x} \in \mathbb{R}^n, \mathbf{x} \in \mathcal{D} \Longrightarrow f( \mathbf{x} ) < 0
\end{equation}
\noindent
$D$ is assumed to be a rectangle given by $D = \{(x_1, \cdots x_n) | a_i \leq x_i \leq b_i \} = [\mathbf{a}, \, \mathbf{b}]$. It is assumed that $f(\mathbf{x})$ is twice continuously differentiable 
in an open domain $U \supset D$. It is proved in \cite{solo} that the inequality deduced from the Taylor expansion of $f$ (\ref{eq:taylor}) allows solving (\ref{eq:main_taylor}).

\begin{equation}\label{eq:taylor}
 f(\mathbf{x}) = f(\mathbf{y}) + \sum_{i = 1}^n \frac{\partial f}{\partial x_i} (\mathbf{y}) (y_i - x_i) + \frac{1}{2} \sum_{i,j = 1}^n \frac{\partial^2 f}{\partial x_i \partial x_j}
 (\mathbf{p}) (y_i - x_i)(y_j - x_j))
\end{equation}

\noindent
where $y \in [\mathbf{a}, \, \mathbf{b}]$, $\mathbf{p} \in [\mathbf{a}, \mathbf{b}]$ and $\mathbf{w} = \max\{ \mathbf{y} - \mathbf{a}, \mathbf{b} - \mathbf{y}  \}$.

Consider a Lyapunov function \cite{Lyap}
$V : \mathbb{R}^n \mapsto \mathbb{R}$ that must verify $V(x) > 0, x \in U - \{ 0\}$ where $U$ is a neighborhood region around $x = 0$; $V(0) = 0$; $\dot{V}(x) \leq 0$, where
$x \in \mathcal{B} - \{ 0 \}$ for some neighborhood close to 0. Such technique can be applied to verify that the condition $\dot{V}(x) \leq 0$.
{\it Supposing that a Lyapunov candidate function is identified for a particular problem of stability, we assume that, according to (\ref{eq:main_taylor}), the formal verification 
of the Lyapunov candidate function consists in solving the problem:

\begin{equation}\label{eq:main_taylor_lypa}
 \forall \mathbf{x} \in \mathbb{R}^n, \mathbf{x} \in \mathcal{D} \Longrightarrow \dot{V}( \mathbf{x} ) < \varepsilon
\end{equation}

\noindent
where $\varepsilon$ is a positive and small constant close to zero.}

\subsection{A basic example}

To illustrate how HOL Light works, consider the linear stable system:

\begin{equation}\label{eq:SS_example}
\frac{d x(t)}{d t} = A x(t) 
\end{equation}
\noindent
where :
\begin{equation}
A = \left( \begin{array}{cc}
-1 & 2 \\
 -3 & -4 \\
 \end{array} \right)
 \end{equation}
 
\noindent
Considering a quadratic Lyapunov metric $\displaystyle{V(x) = \frac{x_1^2 + x_2^2}{2}}$, the derivative of $V$ according to the time verifies:
\begin{equation}\label{eq:lyap_dot}
\dot{V}(x) = - 2x_1(x_1 - 2x_2) - 2x_2(3x_1 + 4x_2)
\end{equation}
Using the tool to verify the nonlinear inequalities \cite{solo}, the following HOL Light script\footnote{The bounds on $x_1$ 
and $x_2$ have been set in $[0, \, 1]$. A further study of the variation of such interval may be considered.} solves the problem considering $\varepsilon = 0.01$:

\begin{verbatim}
let Verify_Lyap_ineq () =
  verify_ineq default_params 5 `&0 <= x /\ x <= &1 /\ &0 <= y /\ y <= &1 
  ==> -- #2.0*x*(x - #2.0*y) - #2.0*y*(#3.0*x + #4.0*y) < #0.01`;;
\end{verbatim}
\noindent
and gives the following solution certificate:

\begin{verbatim}
- : thm * M_verifier_main.verification_stats =
(|- !x y.
        (&0 <= x /\ x <= &1) /\ &0 <= y /\ y <= &1
        ==> --#2.0 * x * (x - #2.0 * y) - #2.0 * y * (#3.0 * x + #4.0 * y) <
            #0.01,
 {total_time = 0.57013392448425293;
  formal_verification_time = 0.0529577732086181641;
  certificate =
   {Verifier.pass = 5; pass_raw = 5; pass_mono = 5; mono = 0; glue = 9;
    glue_convex = 0}})
\end{verbatim}

{\it Verifying (\ref{eq:lyap_dot}) by HOL Light would allow to certify any stability problem relating to dynamical systems} and in particular the Lyapunov candidate 
function built from the B-switched system.

\section{Conclusion and future work}

In this brief, we presented a strategy to describe switched systems as a polynomial system able to mimic the switching behavior using a Bernstein interpolation. 
Such transformation allows "unifying" each switching subsystems into a "single" expression in order to deduce stability properties using a simple Lyapunov function.

Further works includes :
\begin{itemize}
 \item Proving the complete equivalence between the standard switched system representation and the proposed B-switched system representation regarding the stability properties.
 \item Application to nonlinear and non-polynomial systems.
 \item Investigations of the Taylor approximation algorithm in order to not only verify / certify a Lyapunov function, but also calculate the "optimal" Lyapunov function 
 via SOS tools \cite{SOS} \cite{Papa}. 
\end{itemize}
\noindent
The source code of all programs and scripts described in this paper is available upon request to the author.

\section*{Acknowledgement}

The author is sincerely grateful to Dr. Edouard Thomas for his strong guidance and for having introduced to the author, the wonderful world of formal proofs !

\end{document}